\documentclass[aps,prd,amsmath,amssymb,showpacs]{revtex4}

\usepackage{epsfig,float}
\usepackage{graphicx}
\usepackage{dcolumn}% Align table columns on decimal point
\usepackage{morefloats}% bold math
\usepackage{color}
\usepackage{slashed}%\slashed p is ok
\usepackage{bbm}
\usepackage{bm}
\usepackage{mathtools}
\usepackage{booktabs}
\usepackage{multirow}
\usepackage{verbatim}%  \begin{comment} \end{comment}
\allowdisplaybreaks[4]

\newcommand{\la}{\left\langle}
\newcommand{\ra}{\right\rangle}
\newcommand{\rep}[2]{\prescript{#1}{}{#2}}
\newcommand{\eq}{\mathsf{e}}
\newcommand{\Mc}{\mathcal{M}}

\begin{document}

\title{Hyperon Weak Radiative Decay}

\author{Peng-Yu Niu$^{1,2}$\footnote{Email: niupy@ihep.ac.cn}, Jean-Marc Richard$^3$\footnote{E-mail: j-m.richard@ipnl.in2p3.fr}, Qian Wang$^{4}$\footnote{E-mail: qianwang@m.scnu.edu.cn}, Qiang Zhao$^{1,2}$~\footnote{E-mail: zhaoq@ihep.ac.cn, Corresponding author} }

\affiliation{ 1) Institute of High Energy Physics,
        Chinese Academy of Sciences, Beijing 100049, China}

\affiliation{ 2)  University of Chinese Academy of Sciences, Beijing 100049, China}

\affiliation{ 3) Universit\'e de Lyon, Institut de Physique des 2 Infinis de Lyon, UCBL--IN2P3-CNRS,
4, rue Enrico Fermi, Villeurbanne, France}

\affiliation{ 4) Guangdong Provincial Key Laboratory of Nuclear Science,
Institute of Quantum Matter, South China Normal University, Guangzhou 510006, China}

\begin{abstract}

We revisit the hyperon weak radiative decays in the framework of the non-relativistic constituent quark model. This study confirms the nonlocal feature of the hyperon weak radiative transition operators which are dominated by the pole terms, and an overall self-consistent description of the available experimental data for the Cabibbo-favored hyperon weak radiative decays is achieved. It provides a natural mechanism for evading the Hara theorem, where sizeable parity-violating contributions can come from the intermediate orbital excitations. Cancellations between pole terms also explain significant SU(3) flavor symmetry breaking manifested by the experimental data. We also discuss several interesting selection rules arising from either the electromagnetic or the weak interaction vertices. These features suggest nontrivial relations among different hyperon decays.

\end{abstract}
\date{\today}

\maketitle

\section{Introduction}

Although the ground-state hyperons were discovered more than 60 years ago and have played a key role for our understanding of the weak interaction, there are still open questions concerning their weak decay mechanisms. In particular, the hyperon weak radiative decays provide a unique probe for studying the weak, strong and electromagnetic (EM) interaction~\cite{Li:2016tlt}. This process generally has a very small branching ratio and is hard to measure. Recently, BESIII has collected more than 1 billion $J/\psi$ events which can provide a golden opportunity to investigate the properties of the hyperons produced in $J/\psi$ decays. In particular, it is timely to revisit the detailed dynamics of the hyperon weak radiative decays.

One of the long-standing questions associated with the hyperon weak radiative decays is the so-called ``Hara theorem"~\cite{Hara:1964zz}. It was shown in Ref.~\cite{Hara:1964zz} that the parity-violating amplitudes for $\Sigma^+ \to p \gamma $ and $\Xi^- \to \Sigma^- \gamma$ were zero in the limit of unitary symmetry within the pole approximation~\cite{Feldman:1961su}. However, the experimental measurements did not support this prediction and the asymmetry parameter was found large for $\Sigma^+ \to p \gamma$, with a negative sign~\cite{Timm:1994av}.

There is an abundant literature on the theoretical efforts in understanding the physics behind the Hara theorem and on the experimental observations.
A unified theory was proposed by Zenczykowski {\it et al.} \cite{Zenczykowski:1989pt,Zenczykowski:1991mx,Lach:1995we,Zenczykowski:1995ci,Zenczykowski:2001er,Zenczykowski:2002eg,Zenczykowski:2005cs,Zenczykowski:2006se} who combined the SU(6) symmetry with the vector meson dominance for the study of the hyperon weak radiative decays. It was shown in Refs.~\cite{Zenczykowski:2005cs,Zenczykowski:2006se} that  large negative values for the asymmetry parameter in $\Sigma^+\to p \gamma$ were due to the SU(3) flavor symmetry breaking effects. In Ref.~\cite{Close:1980ym} Close and Rubinstein proposed a ``modern pole model" and illustrated the importance of long-distance contributions arising from intermediate pole terms. By estimating the relative SU(6) spin-flavor coupling coefficients they showed the intermediate $1/2^-$ states could have sizeable contributions to the parity-violating amplitude. Various calculations based on the pole dominance scenario in the quark model can be found in the literature~\cite{Picek:1979ky,Gavela:1980bp,Rauh:1981zi,Nardulli:1987ub,Dubovik:2008zz,Chang:2000hu,Bassalleck:1995nz}.  The SU(3) flavor symmetry and pole model are combined to investigate the weak electromagnetic decays of hyperons in Ref.~\cite{Graham:1965zt}. Besides the quark model approaches, chiral perturbation theory (ChPT) has also been applied to the hyperon decays and the intermediate $1/2^-$ states were found to play an essential role in the hyperon radiative decays~\cite{Neufeld:1992np,Jenkins:1992ab,Bos:1996ig,Borasoy:1999nt}. In Ref.~\cite{Balitsky:1989ry} the radiative decay of $\Sigma^+\to p \gamma$ was calculated by an extended QCD sum rule approach. Unitarity and MIT bag model are also employed for the study of hyperon radiative weak decays in the literature~\cite{Farrar:1971xi,Golowich:1982cn}. It is worthy mentioning that in Refs.~\cite{Dubovik:2008zz,Bukina:2001yp} the Hara theorem was shown to  result from the old-fashioned SU(3)$_{\rm f}$ model which cannot avoid the flavor-changing neutral current. In the framework of the Glashow-Iliopoulos-Maiani mechanism, the ``penguin" transition process can evade the flavor-changing neutral current and lead to nonvanishing asymmetry parameter for $\Sigma^+\to p \gamma$~\cite{Shifman:1975tn}. However, such a mechanism seemed not to be sufficient to explain the large value for the asymmetry parameter.

The recognition of the importance of the pole term contributions in the hyperon radiative decays~\cite{Close:1980ym} seems to be crucial for a coherent interpretation of the puzzling experimental data. Similar phenomena have also been found in their hadronic weak decays~\cite{LeYaouanc:1978ef,Richard:2016hac}. A recent investigation of the Cabbibo-favored $\Lambda_c$ hadronic weak decays also showed that the pole terms play a dominant role in the transition amplitudes~\cite{Niu:2020gjw}. There are interesting consequences arising from the pole models. Firstly, it suggests that the hadronic or radiative weak decays are driven by  nonlocal interactions where the strong or radiative interaction and the weak interaction are connected by the intermediate propagators. Secondly, these pole terms can have interferences which leads to large SU(3) flavor symmetry breaking effects. This is understandable since a relatively small SU(3) flavor symmetry breaking in each pole term can get amplified if a destructive interference is involved~\cite{LeYaouanc:1978ef,Richard:2016hac}. Third, the inclusion of the $1/2^-$ states can contribute to large parity-violating effects and thus evade the Hara theorem~\cite{Close:1980ym,Borasoy:1999nt,Gavela:1980bp}. These general points will be addressed in this analysis based on the systematic study of the hyperon radiative decays.

In this work we will revisit the hyperon weak radiative decays in the frame work of the non-relativistic constituent quark model (NRCQM) and provide a coherent description of the Cabbibo-allowed weak radiative decays. To proceed, the details of the framework are presented in Sec.~\ref{sec:frame}. Results and discussions are given in Sec.~\ref{sec:resdis}, and a brief summary is given in Sec.~\ref{sec:sum}. Conventions and analytical amplitudes are provided in Appendix.

\section{Framework}\label{sec:frame}

\begin{figure}[ht]
\begin{center}
\includegraphics[scale=0.6]{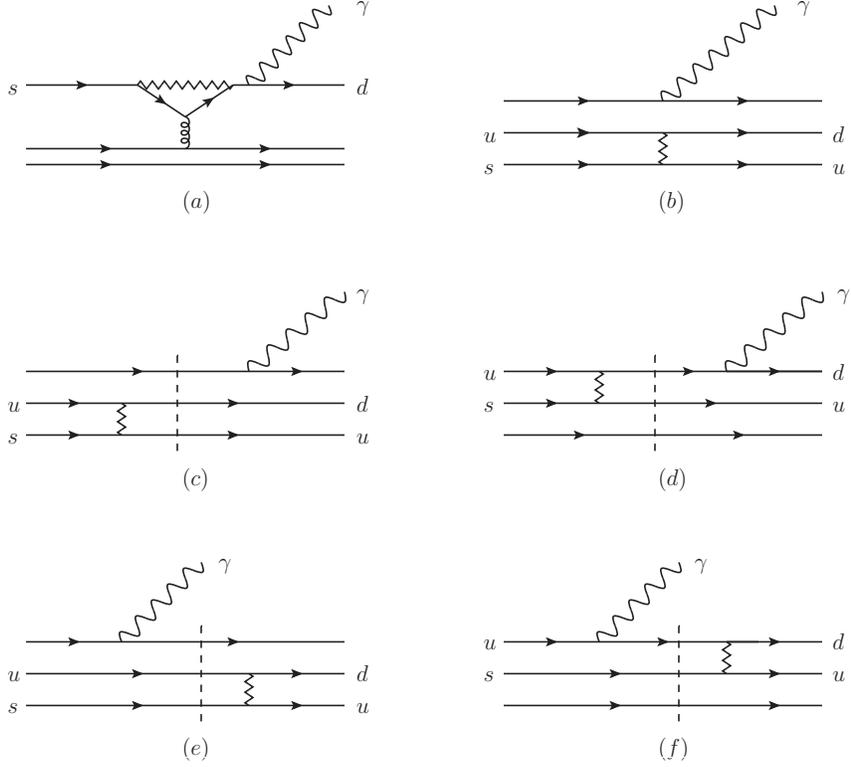}
\caption{Typical weak radiative transitions are categorized: (a) is the typical single-quark transition diagram. (b)-(f) are the two-quark transition diagrams among which (c)-(f) are called pole terms. Note that every diagram actually stands for a group of diagrams, considering that the photon can be emitted by any quark.}
\label{fig:fig1}
\end{center}
\end{figure}

An obvious feature of radiative decays is that the charge is conserved. This suggests that the Cabbibo-allowed weak transition processes occur at leading order either via a single-quark transition through a penguin diagram as shown in Fig.~\ref{fig:fig1}(a), or via a two-quark transition process through an internal conversion of $su\to ud$ as shown in Figs.~\ref{fig:fig1}(b)-(f). The penguin transition is strongly suppressed mainly because it involves loops; this explains the small partial width of $\Xi^-\to \Sigma^-\gamma$. In contrast, the internal conversion processes are tree-level transitions. These are our focus in this work.

The two-quark transition processes can be further categorized into two classes depending on whether intermediate baryons contribute or not, namely, short-distance process and long-distance process. Figure~\ref{fig:fig1}(b) illustrates the short-distance process which stands for a group of diagrams with the photon radiated from any charged particle. Figures~\ref{fig:fig1}(c)-(f) are identified as long-distance processes since there are intermediate resonances (pole terms) contributing in the transition matrix element. For the hyperon weak decays the long-distance pole terms become leading contributions due to the closeness of the intermediate baryons to either the initial or final-state baryons. A coarse estimate of the enhancement factor gives $2\tilde{M}^2/(M_i^2-M_f^2)\simeq \tilde{M}/(M_i-M_f)$, which arises from the intermediate propagators with $\tilde{M}\simeq (M_i+M_f)/2$ a mass scale set by the initial and final-state hyperons. We note in advance that in most cases there exist cancellations among the pole terms of Figs.~\ref{fig:fig1}(c)-(f)~\cite{LeYaouanc:1978ef,Richard:2016hac}. But still the dominance of the pole terms is evident. We thus only consider the pole contributions in this work. Instead of trying to perfectly describe the available data, we intend to appraise the overall quality of the NRCQM approach.

\subsection{Non-relativistic form of the effective Hamiltonian}

Concentrating on the transition amplitudes from the pole terms, the internal conversion via the weak interaction and photon radiation via the EM transition are connected by the intermediate baryons. The weak interaction operator can be described by~\cite{LeYaouanc:1978ef,Close:1980ym,LeYaouanc:1988fx}
\begin{align}
H_W&=\frac{G_F}{\sqrt 2}\int d \bm x \frac12 \{ J^{-,\mu}(\bm x),J^{+}_{\mu}(\bm x) \},
\end{align}
where
\begin{align}
J^{+,\mu}(\bm x)&=\left(\begin{array}{cc}\bar u&\bar c \end{array}\right)\cdot\gamma^\mu(1-\gamma_5)\cdot \left(\begin{array}{cc} \cos \theta_C & \sin \theta_C \notag \\
 -\sin \theta_C &\cos \theta_C \end{array}\right) \cdot\left( \begin{array}{c} d\\s \end{array}\right), \notag  \\
J^{-,\mu}(\bm x)&=\left(\begin{array}{cc}\bar d &\bar c \end{array}\right)\cdot \left(\begin{array}{cc} \cos \theta_C & -\sin \theta_C \\ \sin \theta_C &\cos \theta_C \end{array}\right)\cdot \gamma^\mu(1-\gamma_5)\cdot  \left( \begin{array}{c} u\\c \end{array}\right).
\end{align}
$H_W$ can be separated into two parts with either parity-conserving (PC) or parity-violating (PV) behavior, i.e.,
\begin{equation}
H_{W}\equiv H_W^{PC}+H_W^{PV} ,
\end{equation}
where
\begin{align}
H_W^{PC}&\equiv \frac{G_F}{\sqrt 2} \int \mathrm d \bm x \left[ j_\mu^{(-)}(\bm x)j^{(+)\mu}(\bm x) +j_{5\mu}^{(-)}(\bm x)j_5^{(+)\mu}(\bm x) \right], \\
H_W^{PV}&\equiv \frac{G_F}{\sqrt 2} \int \mathrm d \bm x \left[ j_\mu^{(-)}(\bm x)j_5^{(+)\mu}(\bm x) +j_{5\mu}^{(-)}(\bm x)j^{(+)\mu}(\bm x) \right].
\end{align}
In the non-relativistic limit, $H_W^{PC,PV}$ can be reduced to~\cite{LeYaouanc:1988fx}
\begin{eqnarray}\label{Hamiltonian-PC}
H_W^{PC}&=&\frac{G_F}{\sqrt 2} V_{\text{ud}}V_{\text{us}} \sum_{i\neq j}\frac{1}{(2\pi)^3} \tau^{(-)}_i \nu^{(+)}_j\left( 1- {\bm\sigma}_i\cdot{\bm\sigma}_j\right)\delta(\bm{p}_i^\prime+\bm{p}_j^\prime-\bm{p}_i-\bm{p}_j),
\end{eqnarray}
and
\begin{eqnarray}\label{Hamiltonian-PV}
H_W^{PV}&=&\frac{G_F}{\sqrt 2} V_{\text{ud}}V_{\text{us}} \sum_{i \neq j }\frac{1}{(2\pi)^3}\delta(\bm{p}_i^\prime+\bm{p}_j^\prime-\bm{p}_i-\bm{p}_j) \tau^{(-)}_i \nu^{(+)}_j \nonumber\\
&& \times \frac{1}{2m_q} \left\{ -(\bm{\sigma}_i-\bm{\sigma}_j)\cdot[(\bm{p}_i^\prime-\bm{p}_j^\prime)+(\bm{p}_i-\bm{p}_j)] + i(\bm{\sigma}_i\times\bm{\sigma}_j)\cdot[(\bm{p}_i-\bm{p}_j)-(\bm{p}_i^\prime-\bm{p}_j^\prime)] \right\},
\end{eqnarray}
where $\tau^{(-)}$ and $\nu^{(+)}$ are flavour changing operators which operate as $\tau^{(-)}u=d$ and $\nu^{(+)}s=u$, respectively. The subscript $i, j$ are the quark labels. $m$ is the constituent quark mass. In this work we take the SU(3) flavor symmetry limit with $m_q=m_u=m_d=m_s$. $V_\text{ud}$ and $V_\text{us}$ are the Cabbibo-Kobayashi-Maskawa matrix elements.

The Hamiltonian for the EM interaction is written as:
\begin{align}
H_{EM}=\eq \int d \bm x \bar q(\bm x)\gamma^\mu q(x)A_\mu(\bm x) \ ,
\end{align}
where $\eq$ is the charge of the quark; $q(\bm x)$ and $\bar q(\bm x)$ are the $j^{th}$ quark fields before and after emitting the photon. In the non-relativistic limit, $H_{EM}$ can be expanded in the momentum space as:
\begin{align}
H_{EM}=
&\sum_j \eq_j \bar u(\bm p_j^f)\gamma^\mu u(\bm p_j^i)\epsilon_\mu
\delta(\bm p^f_j+\bm k -\bm p^i_j)\notag\\
=&\frac{1}{(2\pi)^\frac32}\frac{1}{(2k^0)^\frac12}\sum_j \eq_j
\left \{\epsilon_0-\left[\frac{\bm\epsilon \cdot \bm p_j^i}{2m_j}+\frac{\bm\epsilon \cdot \bm p_j^f}{2m_j} +i\frac{\bm\sigma_j \cdot(\bm k\times \bm \epsilon) }{2m_j}\right ] \right\} \delta(\bm p_j^f+\bm k -\bm p_j^i),
\end{align}
where $m_j$ and $\eq_j$ are the mass and charge of the $j^{th}$ quark, respectively; $\bm p^i_j$ and $\bm p^f_j$ denote the three-vector momentum carried by the $j^{th}$ quark before and after emitting the photon; $\bm k$ is the photon momentum and $(\epsilon_0,\bm \epsilon)$ its polarization.  For  real emitted photon, only the transverse polarizations can contribute. $H_{EM}$ can be reduced to:
\begin{align}
\label{eq:EM}
H_{EM}=-\frac{1}{(2\pi)^\frac32}\frac{1}{(2k^0)^\frac12}\sum_j \eq_j
\left[\frac{\bm\epsilon \cdot \bm p^i_j}{m_j}+i\frac{\bm\sigma_j \cdot(\bm k\times \bm\epsilon) }{2m_j}\right]  \delta^3(\bm p_j^f+\bm k -\bm p_j^i).
\end{align}
The first term contributes to the electric-dipole transition which would raise or decrease orbital angular momentum by one unit, and the second term contributes to the magnetic-dipole transition which could raise or decrease the spin of the $j^{th}$ quark by one unit \cite{LeYaouanc:1978ef,Close:1980ym,LeYaouanc:1988fx}.

\subsection{The decay width and asymmetry parameter}

With the operators defined in the previous Subsection we can calculate the pole terms in the NRCQM framework. We take the direction of the photon momentum as the $z$-axis, i.e., $k_0=k$ and $\bm k=(0, \ 0, \ k)$ are the energy and three momentum of photon, respectively. The energy and momentum of initial and final baryon are noted as $(E_i, \ \bm P_i)$ and $(E_f, \ \bm P_f)$, respectively. In the rest frame of the initial baryon, we have $\bm P_i=0$ and $\bm k=-\bm P_f$.

The nonlocal operators for the weak and EM transitions will distinguish processes between Figs.~\ref{fig:fig1}(c) and (e) (or between Figs.~\ref{fig:fig1}(d) and (f)). With the baryon wavefunctions constructed on the basis of SU(6)$\otimes$O(3) symmetry and by separately calculating the weak internal conversion and EM transition couplings, Figs.~\ref{fig:fig1}(c) and (e) together will be evaluated explicitly with the quark labels tagged to the interacting quarks, and similarly for Figs.~\ref{fig:fig1}(d) and (f). Note that in the literature, e.g.,~\cite{Dubovik:2008zz}, Fig.~\ref{fig:fig1}(c) and (d) (or Figs.~\ref{fig:fig1}(e) and (f)) are sometimes treated as two different transition processes according to  whether or not the EM transition operator is involved in the weak internal conversion process. If the symmetry is properly accounted for both wavefunctions and operators, in principle there is no need to distinguish between Figs.~\ref{fig:fig1}(c) and (d), or between (e) and (f). For convenience, we label the amplitudes of Figs.~\ref{fig:fig1}(c)-(d) and Fig.~\ref{fig:fig1}(e)-(f) by the subscript $A$ and $B$, respectively.

Proceeding to the calculation of the full transition amplitudes, we separate the parity-conserving and parity-violating parts as follows:
\begin{align}
\Mc=\Mc(\bm \epsilon,S_f^z,S_i^z)=\Mc_{PC}+\Mc_{PV},
\end{align}
where $S_f^z$ and $S_i^z$ are the third component of spin of final baryon and initial baryon, respectively, and are omitted to lighten the writting. Also omitted is the photon polarization $\bm \epsilon=\mp (1,\pm i,0)/\sqrt2$. Thanks to the symmetries and hermiticity of the Hamiltonian, the amplitudes corresponding to the two possible polarizations are related, and it is sufficient to do the calculation with the polarization  $\bm \epsilon=-(1,i,0)/\sqrt2$, abbreviated as ''+``. The parity-conserving amplitude $\Mc_{PC}$ and parity-violating amplitude $\Mc_{PV}$ are then given by the pole terms via the process A or B, i.e.,
\begin{align}
\Mc_{PC}&=\Mc_{PC,A}+ \Mc_{PC,B},\\
\Mc_{PV}&=\Mc_{PV,A}+ \Mc_{PV,B},
\end{align}
where
\begin{align}
\Mc_{PC,A}\equiv \sum _{B_m} \la B_f(\bm P_f,S_f^z) \middle| H_{EM} \middle| B_m(\bm P_i,S_i^z) \ra \frac{i}{\slashed p_{B_m} -m_{B_m}}\la B_m(\bm P_i,S_i^z) \middle| H_W^{PC} \middle| B_i(\bm P_i,S_i^z) \ra, \\
\Mc_{PV,A}\equiv \sum _{B'_m} \la B_f(\bm P_f,S_f^z) \middle| H_{EM} \middle| B'_m(\bm P_i,S_i^z) \ra \frac{i}{\slashed p_{B'_m} -m_{B'_m}}\la B'_m(\bm P_i,S_i^z) \middle| H_W^{PV} \middle| B_i(\bm P_i,S_i^z) \ra, \\
\Mc_{PC,B}\equiv \sum _{B_m} \la B_f(\bm P_f,S_f^z) \middle| H_{W}^{PC} \middle| B_m(\bm P_f,S_f^z) \ra \frac{i}{\slashed p_{B_m} -m_{B_m}}\la B_m(\bm P_f,S_f^z) \middle| H_{EM} \middle| B_i(\bm P_i,S_i^z) \ra, \\
\Mc_{PV,B}\equiv \sum _{B'_m} \la B_f(\bm P_f,S_f^z) \middle| H_{W}^{PV} \middle| B'_m(\bm P_f,S_f^z) \ra \frac{i}{\slashed p_{B'_m} -m_{B'_m}}\la B'_m(\bm P_f,S_f^z) \middle| H_{EM} \middle| B_i(\bm P_i,S_i^z) \ra.
\end{align}
where a complete set of intermediate baryon states $B_m$ ($B_m'$) with quantum numbers $1/2^+$ ($1/2^-$) has been included in process A and B, respectively.  $\langle B_m | H_{W}^{PC/PV} | B_m'\rangle$ and $\langle B_m| H_{EM} | B_m' \rangle$ are the weak and EM transition matrix elements, respectively.  The weak matrix elements are calculated in a similar way as in Refs.~\cite{Richard:2016hac,Niu:2020gjw}. In the following we provide some details about the calculations of the EM transition matrix elements.

Taking advantage that we use baryon wave functions that are fully symmetric with respect  to the space, spin and flavor degrees of freedom (see Appendix), $H_{EM}$ can be replaced as
\begin{align}
H^{+}_{EM}&= \sum_j (H_{EM}^+)_j \to 3(H_{EM}^+)_1 =-\frac{1}{(2\pi)^\frac32}\frac{1}{(2k^0)^\frac12}\frac{3\eq_1}{\sqrt2 m_1}
\left[p_1^+ +\sigma_1^+ k \right]  \delta^3(\bm p_1^f+\bm k -\bm p_1^i).
\end{align}
where
\begin{align}
p^+\equiv p_x+ip_y, ~~~\sigma^+\equiv \frac{\sigma_x+i\sigma_y}{2}.
\end{align}

Taking the decay of $\Lambda\to \gamma n$ of $A$-type as an example, the EM transition amplitude can be obtained as
\begin{align}\label{EM-operator-1}
&\la B_f(\bm P_f,S_f^z)\middle | H_{EM}^+ \middle | B_i(\bm P_i,S_i^z) \ra  \notag \\
&{}= \la \frac{1}{\sqrt2}(\phi_n^\rho\chi_{1/2,S_f^z}^\rho+\phi_n^\lambda\chi_{1/2,S_f^z}^\lambda)\Phi^{00;00}_{00}(\bm p_1^f,\bm p_2^f,\bm p_3^f) \middle| H_{EM}^+ \middle | \frac{1}{\sqrt2}(\phi_n^\rho\chi_{1/2,S_i^z}^\rho+\phi_n^\lambda\chi_{1/2,S_i^z}^\lambda)\Phi^{00;00}_{00}(\bm p_1^i,\bm p_2^i,\bm p_3^i) \ra \notag \\
&{}={}C \langle\Phi^{00;00}_{00}(\{\bm p_j^f\})|p^+_1|\Phi^{00;00}_{00}(\{\bm p_j^i\})\rangle+ D \langle\Phi^{00;00}_{00}(\{\bm p_j^f\})|k|\Phi^{00;00}_{00}(\{\bm p_j^i\})\rangle.
\end{align}
Here $\Phi(\bm p_1,\bm p_2,\bm p_3)=\Psi(\bm P,\bm p_\rho,\bm p_\lambda)$ is the wave function written in terms of individual momenta and spelled out in the Appendix as a function of the Jacobi coordinates, and its indices denote both the internal and global radial and orbital excitations. Accordingly, the two last brackets
are the matrix elements of these spatial wave functions in the momentum space, and the coefficients $C$ and $D$ are the factors extracted in the spin-isospin space. The first term will raise the orbital angular momentum projection of the interacting quark by one unit and the second one will raise its spin projection by one unit.
A general definition of the convolution integral reads
\begin{align}\label{integral-def}
\langle \cdots O(p) \cdots \rangle&{}=I_{n_\rho^f \ell_\rho^f;n_\lambda^f\ell_\lambda^f;L^f M^f;n_\rho^i \ell_\rho^i;n_\lambda^i \ell_\lambda^i;L^i M^i}(O(p))\notag\\
&{}= \int d^{(3)}\bm p^f d^{(3)}\bm p^i \delta^3(\bm p_1^f+\bm k -\bm p^i_1)\delta^3(\bm p_2^f-\bm p_2^i)\delta^3(\bm p_3^f-\bm p_3^i)\notag\\
&{} \quad\times \Phi^*{\,}_{L^f M^f}^{n_\rho^f\ell_\rho^f;n_\lambda^f{}\ell_\lambda^f}(\bm p_1^f,\bm p_2^f,\bm p_3^f) O(p)
\Phi_{L^i M^i}^{n_\rho^i\ell_\rho^i;n_\lambda^i{}\ell_\lambda^i}(\bm p_1^i,\bm p_2^i,\bm p_3^i),
\end{align}
where
$O(p)$ is a  function of quark momenta, such as $p_1^+$ and $k$.
For the transitions between two ground states, it is easy to verify that
\begin{align}
I(p^+_1)=0,\qquad I(k)= k e^{-\genfrac{}{}{}{1}{k^2}{6 \alpha ^2}}.
\end{align}
This indicates that only the term proportional to $D$, namely, the magnetic-dipole part of $H^+_{EM}$ in Eq.~(\ref{EM-operator-1}), can contribute to the transitions between the two ground states. With fixed spin projections in the initial and final states, i.e., $S_i^z=-1/2$ and $S_f^z=1/2$, coefficient $D$ can be calculated in the spin-flavor space
\begin{align}
D&=-\frac{3}{(2\pi)^\frac32}\frac{1}{(2k^0)^\frac12}
\la \frac{1}{\sqrt2}\left(\phi_n^\rho\chi_{1/2,1/2}^\rho+\phi_n^\lambda\chi_{1/2,1/2}^\lambda \right)
\middle|\frac{\eq_1 \sigma^+_1}{\sqrt2 m_1}\middle |
\frac{1}{\sqrt2}\left(\phi_n^\rho\chi_{1/2,-1/2}^\rho+\phi_n^\lambda\chi_{1/2,-1/2}^\lambda \right)\ra \notag \\
&=\frac{\eq}{6 \sqrt{2} \pi ^{3/2} \sqrt{k_0} m_q}.
\end{align}
Finally, we have
\begin{align}
\la n \left(\bm P_f,\frac12 \right) \middle | H_{EM}^+ \middle | n\left(\bm P_i,-\frac12\right)\ra=\frac{\eq k}{6 \sqrt{2} \pi^{3/2} \sqrt{k_0} m_q} e^{-\genfrac{}{}{}{1}{k^2}{6 \alpha ^2}}.
\end{align}
Analogously, the radiative transitions between the intermediate $1/2^-$ baryon and final $1/2^+$ baryon can be calculated. In such processes the contributions will come from the term proportional to $C$ in Eq.~(\ref{EM-operator-1}). The wavefunctions and detailed expressions of the transition amplitudes are provided in the Appendix.

With the explicit amplitudes for the PC and PV transitions, the partial decay width for the hyperon weak radiative decays can be obtained as follows:
\begin{align}
\Gamma&=8\pi^2 \frac{|\mathbf k| k_0 E_f}{M_i}\frac{1}{2S_i+1}\sum_{S_f^z,S_i^z} 2\left(|\Mc^{+,S_f^z,S_i^z}_{PC}|^2 + |\Mc^{+,S_f^z,S_i^z}_{PV}|^2\right),
\end{align}
where $\Mc_{PV}(+,S_f^z,S_i^z)$ and $\Mc_{PV}(+,S_f^z,S_i^z)$ are the PC and PV amplitudes, respectively, with the photon polarization $\bm \epsilon ^+$, and $S_i$ and $M_i$ are the spin and mass of the initial baryon, respectively.

The parity asymmetry parameter can be extracted in the quark model.
Generally, the amplitude of hyperon weak radiative decay at hadron level has the following form:
\begin{align}
\label{eq:AH}
\Mc=G_F \frac{\eq}{\sqrt{4\pi}}\epsilon_\mu \bar u(p')(A+B\gamma_5)\sigma^{\mu\nu}q_\nu u(p),
\end{align}
where $\epsilon_\mu$ is the polarization vector of the photon; $u(p)$ and $\bar u(p')$ are the spinors of the initial-state hyperon and final-state nonstrange baryon, respectively. $A$ and $B$ represent the PC ($P$-wave) and PV ($S$-wave) amplitudes, respectively. The asymmetry parameter is defined as:
\begin{align}
\label{eq:asp}
\alpha_\gamma\equiv \frac{2 \text{Re}(A^*B)}{|A|^2+|B|^2}.
\end{align}
By matching $A$ and $B$ to the quark model amplitudes,  the asymmetry parameter $\alpha_\gamma$ can be expressed in terms of the quark model amplitudes for the PC and PV transitions, i.e.,
\begin{align}
\label{eq:asp2}
\alpha_\gamma=\frac{2 \text{Re}(\Mc_{PC}^*\Mc_{PV})}{|\Mc_{PC}|^2+|\Mc_{PV}|^2}.
\end{align}
With this asymmetry parameter, the angular distribution of the final-state baryon in the rest frame of the initial hyperon can be written as:
\begin{align}
\frac{d N}{d \Omega}=\frac{N}{4\pi}\left (1+\alpha_\gamma  \bm P_h \cdot \hat{\bm p}  \right ),
\end{align}
where $\bm P_h$ is the polarization vector of the decaying hyperon and $\hat{\bm p}$ is the direction of the three-vector momentum of the final-state baryon.

The following processes are included in this work: $\Lambda \to n \gamma$, $\Sigma^+\to p\gamma$, $\Sigma^0\to n\gamma$, $\Xi^0 \to \Lambda\gamma$ and $\Xi^0 \to \Sigma^0 \gamma$.
Although the partial width of $\Sigma^0\to \Lambda \gamma$ is saturated by the EM interaction, we still evaluate the weak radiative decay contribution here as a comparison with the EM process.

\section{Results and Discussions}\label{sec:resdis}

For these weak radiative decay processes, i.e., $\Lambda \to n \gamma$, $\Sigma^+ \to p \gamma$, $\Sigma^0 \to n \gamma$, $\Xi^0 \to \Lambda\gamma$ and $\Xi^0 \to \Sigma^0\gamma$, they are all Cabbibo-favored with $|\Delta I|=1/2$ and $|\Delta s|=1$ transition. The dominance of the pole terms in the transition amplitudes suggests that all the intermediate states of $J^P=1/2^+$ with the proper flavor should be included for the PC amplitude, while all the states of $1/2^-$ should be included for the PV one. However, taking into account the propagator suppression effects when the intermediate states become highly off-shell, we only consider the first orbital excitation states for the PV amplitude.  This approximation will bring some uncertainties to the final results. Taking the $\Lambda$ decays as an example, the upper limit of the uncertainties can be estimated by the ratios $R_{PC}=|(M_\Lambda^2-M_p^2)/(M_\Lambda^2-M_{N^*(1440)}^2)|^2\simeq 0.20$ for the PC part in the branching ratio, and $R_{PV}=|(M_\Lambda(1405)^2-M_p^2)/(M_{\Lambda(L=3)}^2-M_p^2)|^2\simeq 0.12$ for the PV part in the branching ratio. Here, $N^*(1440)$ is the first radial excitation state of the nucleon and $\Lambda(L=3)$ denotes the second negative parity orbital excitation state of $\Lambda$. The multiplets with $L=3$ have not yet established in experiment and we adopt $M_{\Lambda(L=3)}=2$ GeV for its mass as a conservative estimate. Theoretical calculations in the literature suggest that their masses are well above 2 GeV~\cite{Isgur:1978xj} (see also Ref.~\cite{Capstick:2000qj} for a review of baryon spectroscopy in the quark model and references therein).

The intermediate states considered in this work are listed in Tab.~\ref{tab:channel}. These states are either ground states with $J^P=1/2^+$ or first orbital excitations with $J^P=1/2^-$. For the $1/2^-$ intermediate hyperons, $\Lambda(1405)$ is assigned as the flavor singlet in the representation $[70,\rep21]$. For multiplets of representations $[70,\rep28]$ and $[70,\rep48]$, the corresponding states have not yet been determined in experiment. Although the quantum numbers of $\Xi(1620)$ and $\Xi(1690)$ have not been measured in experiment, they adopt the assignments that they belong to representations $[70,\rep28]$ and $[70,\rep48]$, respectively.

\begin{table}[htbp]
\centering
\caption{The intermediate states considered in our calculation. The baryons masses and decay widths (given in the brackets) are taken from PDG~\cite{Olive:2016xmw} (in units of GeV). Only the central values of the masses and widths are listed.}
\begin{ruledtabular}
\begin{tabular}{cccccc}
%\hline \hline
\multirow{4}[0]{*}{PC}& \multirow{4}[0]{*}{$[56,\rep28]$}
  &$p$          &$n$          &$\Lambda$  & $\Xi^0$    \\
& &$0.94$       &$0.94$       &$1.12$     & $1.31$     \\[2pt]
& &$\Sigma^+$   & $\Sigma^0$  &           &            \\
& &$1.20$       & $1.20$      &           &            \\
\hline
\multirow{6}[0]{*}{PV} & \multirow{2}[0]{*}{$[70,\rep28]$}
&$N(1535)$   &$\Lambda(1670)$&$\Sigma(1620)$&$\Xi(1620)$  \\
& &$1.53(0.15)$&$1.67(0.035)$  &$1.62(0.05)$   &$1.62(0.03)$  \\[2pt]
 & \multirow{2}[0]{*}{$[70,\rep48]$}
&$N(1650)$      &$\Lambda(1800)$&$\Sigma(1750)$&$\Xi(1690)$  \\
& &$1.655(0.135)$ &$1.8(0.3)$     &$1.75(0.15)$  &$1.62(0.03)$  \\[2pt]
&\multirow{2}[0]{*}{$[70,\rep{2}{10}]$}
&$\Delta(1620)$ &  &     &\\
& &1.61(0.13)   &  &     &\\
%\hline \hline
\end{tabular}\end{ruledtabular}
\label{tab:mass}
\end{table}

With the EM and weak coupling matrix elements for the $1/2^+$ and $1/2^-$ states calculated in the NRCQM we can obtain the analytical amplitudes for each pole term and they are given in Appendix~\ref{app:amp}. One can see that the PC amplitudes are given by the intermediate $1/2^+$ octet baryons and the decays are through a $P$ wave, while the parity-violating ones are through an $S$ wave.

In our frame work, the input parameters include the constituent quark masses and harmonic oscillator strengths. In the present calculations we
take the SU(3) flavor symmetry for the constituent quark masses, i.e. $m_q=m_u=m_d=m_s=0.35$ GeV, as the leading order approximation. This simplifies the baryon wavefunctions at this moment. Since the three-vector momenta carried by the photon is rather small this approximation does not cause significant uncertainties to the numerical results. In contrast, the numerical results are more sensitive to the harmonic oscillator strengths, i.e. $\alpha_\rho$ and $\alpha_\lambda$ for the $\rho$ and $\lambda$ degrees of freedom in the Jacobi coordinate. In the equal mass limit of $m_u=m_d=m_s$ these two parameters satisfy $\alpha_\rho=\alpha_\lambda\equiv \alpha$. One can see later that the analytical amplitudes will be proportional to $\alpha^3$. Thus, more uncertainties can arise from the harmonic oscillator strength $\alpha$. Considering that $\Xi$ contains two $s$ quarks, the SU(3) flavor symmetry breaking effect should not be ignored, we take a different value for the oscillator strength for the $\Xi$ decays. Thus, in the numerical studies, we adopt $\alpha=0.45$ GeV for the $\Lambda$ and $\Sigma$ decays and $\beta=0.38$ GeV for the $\Xi$ decays.

\begin{table}[htbp]
  \centering
  \caption{The numerical results of amplitudes for every channel, in units of  $(10^{-10}~\text{GeV}^{-1/2})$. }
\begin{ruledtabular}
    \begin{tabular}{ccccccccccc}
%\hline \hline
process & \multicolumn{2}{c}{PCA} & \multicolumn{2}{c}{PVA} & \multicolumn{2}{c}{PCB} & \multicolumn{2}{c}{PVB} & \multicolumn{1}{l}{Total(PC)} &\multicolumn{1}{l}{Total(PV)} \\
\hline
\multirow{5}[0]{*}{$\Lambda \to n\gamma$}
&$n$ &$-6.68$ &$N(1535)$ &$-4.72-0.99i$ &$\Lambda$ &$3.97$ &$\Lambda(1670)$ &$-1.54-0.047i$
& \multirow{5}[0]{*}{$5.82$} & \multirow{5}[0]{*}{$-14.91-1.48i$} \\
&     &    & $N(1650)$ & $0.22+0.032i$ & $\Sigma^0$ & $8.53$  & $\Lambda(1800)$ & $0.072-0.017i$ &      &  \\
&     &    & $\Delta(1620)$  &$0$      &       &       & $\Lambda(1405)$ & $-4.27-0.28i$ &       &  \\
&     &    &    &      &       &       & $\Sigma(1620)$ & $-4.91-0.23i$   &       &  \\
&     &    &    &      &       &       & $\Sigma(1750)$ & $0.24$ &       &  \\
\hline
\multirow{2}[0]{*}{$\Sigma^+\to p\gamma$}
&$p$ &$-19.06$ &$N(1535)$ &$-9.65-2.39i$ &$\Sigma^+$ &$24.16$ &$\Sigma(1620)$ &$-5.75-0.27i$ &\multirow{2}[0]{*}{$5.10$} & \multirow{3}[0]{*}{$-15.40-2.66i$} \\
&     &     & $N(1650)$ & $0$     &     &     & $\Sigma(1750)$ & $0$ &       &  \\
&     &     & $\Delta(1620)$ &$0$   &     &     &                &     &       &  \\
\hline
\multirow{5}[0]{*}{$\Sigma^0\to n\gamma$}
&$n$ &$8.93$ &$N(1535)$ &$7.69+1.92i$ &$\Lambda$ &$-7.91$ &$\Lambda(1670)$ & $2.15+0.066i$
&\multirow{5}[0]{*}{$6.69$} & \multirow{5}[0]{*}{$1.33+1.37i$} \\
&   &   & $N(1650)$ & $-0.68-0.15i$ & $\Sigma^0$ & $5.67$  &$\Lambda(1800)$ & $-0.20-0.046i$ &       &  \\
&   &   & $\Delta(1620)$ &$0$       &       &       & $\Lambda(1405)$& $-5.58-0.36i$ &       &  \\
&   &   &     &       &       &       & $\Sigma(1620)$ & $-2.28-0.11i$ &       &  \\
&   &   &     &       &       &       & $\Sigma(1750)$ & $0.22$ &       &  \\
\hline
\multirow{5}[0]{*}{$\Xi^0\to \Lambda\gamma$}
&$\Lambda$ &$5.79$ &$\Lambda(1670)$ &$4.73+0.26i$ &$\Xi^0$ &$-13.60$ &$\Xi(1620)$ & $6.33+0.22i$
&\multirow{5}[0]{*}{$-7.81$} &\multirow{5}[0]{*}{$-4.38-3.88i$} \\
& $\Sigma^0$   &$0$  &$\Lambda(1800)$ &$-0.23-0.083i$ &    &   &$\Xi(1690)$ &$-0.46-0.014i$ &    &  \\
&   &   & $\Lambda(1405)$ & $-14.7-4.26i$ &   &   &      &       &       &  \\
&   &   & $\Sigma(1620)$  &$0$              &  &   &  &       &       &   \\
&   &   & $\Sigma(1750)$  &$0$              &  &   &  &       &       &   \\
\hline
\multirow{5}[0]{*}{$\Xi^0\to \Sigma^0\gamma$}
&$\Lambda$ &$-8.15$ &$\Lambda(1670)$ &$-10.75-0.59i$ &$\Xi^0$ &$0$     &$\Xi(1620)$ &$0$     &\multirow{5}[0]{*}{$-8.15$} & \multirow{5}[0]{*}{$-45.65-10.67i$} \\
&$\Sigma^0$   &$0$   & $\Lambda(1800)$ & $0.21+0.074i$ &  &   & $\Xi(1690)$ & $0$   &       &  \\
&   &   & $\Lambda(1405)$ & $-35.11-10.15i$ &  &   &  &       &       &  \\
&   &   & $\Sigma(1620)$  &$0$              &  &   &  &       &       &   \\
&   &   & $\Sigma(1750)$  &$0$              &  &   &  &       &       &   \\
%\hline\hline
    \end{tabular}
    \end{ruledtabular}
  \label{tab:channel}
\end{table}

The numerical results of amplitudes are given in Tab.~\ref{tab:channel}. Some general features can be learned as follows.

\subsubsection{ $\Lambda \to n \gamma$, $\Sigma^+\to p\gamma$, and $\Sigma^0\to n\gamma$}

For the processes $\Sigma^+\to p\gamma$ and $\Lambda \to n \gamma$, it appears that the processes of $A$ and $B$-type have a destructive interference in the the PC channels, but become constructive in the PV channels. This results in that the value of the PC amplitudes is smaller than the value of the PV ones for these two decays. Furthermore, this leads to a large value for the asymmetry parameter.

It is interesting to note that the PV amplitude of the $N(1650)$ which is assigned as the SU(6) representation $[70,\rep48]$ with $J^P=1/2^-$ vanishes in $\Sigma^+\to p\gamma$, but contributes to the neutral channels of $\Lambda \to n \gamma$ and $\Sigma^0\to n\gamma$. This is understandable via the EM coupling between $N^*$ of $[70,\rep48]$ and proton $[56,\rep28]$. In such an EM transition process, the so-called ``Moorhouse selection rule"~\cite{Moorhouse:1966jn} should play a role which prevents the intermediate $N(1650)$ decays into $p\gamma$.

To be more specific, one can prove that the $\rho$ mode decays or excitations between states of $[70,\rep48]$ and $[56,\rep28]$ via the EM transition operator vanishes. Since the spin and orbital angular momentum operators are symmetric to the quark indices of the first two quarks, the $\rho$ mode spatial wavefunction of $[70,\rep48]$ cannot orbitally decay into a symmetric spatial wavefunction of $[56,\rep28]$. The $\lambda$ mode decays or excitations between states of $[70,\rep48]$ and $[56,\rep28]$ are, in principle, allowed. However, for the charged channel of $N^*([70,\rep48])\to p\gamma$ one has $\langle \phi^\lambda_{N^{*+}}|\Sigma_{i=1}^3 \eq_i|\phi^\lambda_p\rangle=3\langle \phi^\lambda_{N^{*+}}|\eq_3|\phi^\lambda_p\rangle=0$. Thus, the $\lambda$ mode is also forbidden.

It should be noted that representations of $[70,\rep48]$ and $[70,\rep28]$ in the first orbital excitations can mix and the physical states $N(1535)$ and $N(1650)$ are actually mixing states of these two configurations. As a consequence, both states can actually contribute in $\Sigma^+\to p\gamma$. This is an interesting phenomenon which may provide an alternative way to study the structure of these two resonances.

Similar relation also appears in the EM transitions of the PV $B$-type process, i.e. the EM transitions between $\Sigma^+$ of $[56,\rep28]$ and $\Sigma^{*+}$ of $[70,\rep48]$. Again, one can prove that only the $\lambda$ mode of transitions is allowed. However, with $\langle \phi^\lambda_{\Sigma^{*+}}|\Sigma_{i=1}^3 \eq_i|\phi^\lambda_{\Sigma^+}\rangle=3\langle \phi^\lambda_{\Sigma^{*+}}|\eq_3|\phi^\lambda_{\Sigma^+}\rangle=0$, the $\lambda$ mode of transitions will also be forbidden. Thus, states of $[70,\rep48]$ do not contribute to the PV amplitude in $\Sigma^+\to p\gamma$.

One also notices that the states of $[70,\rep{2}{10}]$ do not contribute to the octet radiative weak decays. This is due to requirement of the $\rho$ mode orbital excitations between the spatial wavefunctions of $[70,\rep{2}{10}]$ and $[56,\rep28]$ in association with the  spin-flavor operators in Eq.~(\ref{Hamiltonian-PV}) which are nonvanishing between $\lambda$ and $\rho$ modes. As a consequence, the two terms in Eq.~(\ref{Hamiltonian-PV}) exactly cancel out. This vanishing transition was listed in Ref.~\cite{Gavela:1980bp}, but was not discussed much there. Note that the transition of $\Sigma^+\to p\gamma$ does not violate the $|\Delta I|=1/2$ rule. The vanishing contribution of the $\Delta(1620)$ of representation $[70,\rep{2}{10}]$ can be regarded as kind of dynamical selection rule in the quark model. In fact, this selection rule also plays a role in the hadronic weak decays of $\Sigma\to N\pi$ and $\Lambda\to N\pi$. A recent study can be found in Ref.~\cite{Niu:2020gjw}.

\subsubsection{$\Xi^0 \to \Sigma^0 \gamma$ and $\Lambda \gamma$}

The situation in  $\Xi^0\to \Sigma^0 \gamma$ and $\Lambda \gamma$ is different from that in $\Lambda$ and $\Sigma$ decays. Here, the contributions from the PC processes are lager than the destructive PV ones.

One notices that the $B$-type amplitudes in $\Xi^0 \to \Sigma^0 \gamma$ vanish for both PC and PV transitions. Meanwhile, the intermediate $\Sigma^{(*)}$ states do not contribute to the $A$-type transitions.  These are due to the weak interaction operators. Note that the flavor transition element vanishes: $\langle \phi_{\Sigma^0}^\rho|\tau_1^{(-)}\nu_2^{(+)}|\phi_{\Xi^0}^\rho\rangle=0$. In the PC $B$-type transition it leaves the $\lambda$ mode to contribute. However, one finds that the spin transition element $\langle\chi_{\frac 12 S_z}^\lambda|(1-{\bm\sigma}_1\cdot{\bm\sigma}_2)|\chi_{\frac 12 S_z}^\lambda\rangle=0$ with the spins of the first two quarks in parallel. Therefore, the intermediate $\Sigma^0$ does not contribute to the PC amplitude.

For the PV $A$-type transitions the intermediate $\Sigma^{*0}$ belongs to either representation $[70,\rep28]$ or $[70,\rep48]$. Again, with $\langle \phi_{\Sigma^0}^\rho|\tau_1^{(-)}\nu_2^{(+)}|\phi_{\Xi^0}^\rho\rangle=0$ it leaves the wavefunction component $\phi_{\Sigma^0}^\lambda(\chi_{\frac 12 S_z}^\rho\psi_{11L_z}^\rho({\bm\rho}, {\bm\lambda})-\chi_{\frac 12 S_z}^\lambda\psi_{11L_z}^\lambda({\bm\rho}, {\bm\lambda}))$ of $[70,\rep28]$ to be considered. From the PV operators in Eq.~(\ref{Hamiltonian-PV}) one can easily prove that only the transitions between $\rho$ and $\lambda$ mode in the spin and spatial spaces can survive. However, the two terms in Eq.~(\ref{Hamiltonian-PV}) have the same average values but opposite signs. They will cancel and lead to vanishing contributions from the intermediate $\Sigma^{*0}$ of $[70,\rep28]$. For the intermediate $\Sigma^{*0}$ of $[70,\rep48]$ one finds that the corresponding spatial transitions between the $\rho$ mode will vanish.

For the $B$-type transitions the intermediate $\Xi^{(*)0}$ can be in either $[56,\rep28]$ (PC) or $[70,\rep28]$ (PV). Their transitions into the final $\Sigma^0$ vanish for the same reason arising from the weak transition operators.  Thus, we have the interesting results that all the $\Xi^{(*)0}$ pole terms in the $B$-type processes and all the $\Sigma^0$ pole terms in the $A$-type ones vanish in $\Xi^0 \to \Sigma^0 \gamma$.

\subsubsection{Branching ratios and asymmetry parameters}

\begin{table}[ht]
\centering
\caption{The calculated branching ratios (in units of $10^{-3}$) are compared with experimental data and other theoretical predictions. In Ref.~\cite{Borasoy:1999nt} the results are given by decay width.}
\begin{ruledtabular}
\begin{tabular}{lccccccc}
%\hline \hline
$B_i\to B_f \gamma$   &PDG data~\cite{Olive:2016xmw} &Broken SU(3)~\cite{Zenczykowski:2005cs} &ChPT~\cite{Borasoy:1999nt} &Pole model~\cite{Nardulli:1987ub}&Pole model~\cite{Gavela:1980bp} &Our result  \\
\hline
$\Sigma^+\to p\gamma$   &$(1.23\pm0.05)$&$0.72$&$\approx 16$  &$0.75\pm0.30$&$1.15$ &$1.06\pm0.59$       \\
$\Lambda\to n\gamma$    &$(1.75\pm0.15)$&$0.77$&$\approx 1.45$&$0.16\pm0.06$&$0.62$ &$1.83\pm0.96$      \\
$\Xi^0\to\Lambda\gamma$ &$(1.17\pm0.07)$&$1.02$&$\approx 1.17$&$0.72\pm0.42$&$3.0$  &$0.96\pm0.32$       \\
$\Xi^0\to\Sigma^0\gamma$&$3.33\pm0.1$   &$4.42$&$\approx 1.14$&$2.6\pm1.2$  &$7.2$  &$9.75\pm4.18$        \\
$\Sigma^0\to n\gamma$   &-              &-     &$\approx 10^{-9}$&$1.8\times10^{-9}$&$10^{-10}$  &$\approx 10^{-10}$\\
%\hline \hline
\end{tabular}
\end{ruledtabular}
\label{tab:br}
\end{table}

Proceeding to the calculations of experimental observables, the calculated branching ratios and parity asymmetry parameters are listed in the last columns of Tab.~\ref{tab:br} and ~\ref{tab:as}, respectively, in comparison with other models. As shown in Tab.~\ref{tab:br} the central values of the branching ratios are close to the experimental results except for $\Xi^0\to \Sigma^0 \gamma$. By introducing 5\% errors to the quark model parameters, i.e. $\alpha,~\beta$ and $m_q$, we estimate the uncertainties of our model calculations.  It shows that the uncertainties with the branching ratios are around $ 40\% \sim 50\% $ for each channel which means the amplitudes are sensitive to the quark model parameters. This is understandable since the amplitudes are proportional to $\alpha^3/m_q$ ($\beta^3/m_q$).
In contrast, the uncertainties with the asymmetry parameter is relatively smaller. As seen from the expression of Eq.~\eqref{eq:asp2} of $\alpha_\gamma$, the uncertainties arising from the dependence of $\alpha^3/m_q$ or $\beta^3/m_q$ are largely  cancelled out.

Note that there are only a limited number of parameters under the NRCQM framework. The overall quality of our model calculations turns out to be reasonable. Our results indicate that the lowest lying states play an essential role through the intermediate pole terms which is consistent with the result of ChPT~\cite{Borasoy:1999nt} and previous quark model calculations~\cite{Gavela:1980bp}.

In Tab.~\ref{tab:as} one can see that the signs of the asymmetry parameters are quite different.
Our model gives the right signs for the $\Sigma^+$ and $\Lambda$ channel and prefers a large value. However, for the $\Xi^0$ channels the sign of the asymmetry parameter is opposite with the experimental data. The sign of the asymmetry parameter of $\Sigma^0\to n \gamma$ is positive which is opposite to the other model calculations~\cite{Nardulli:1987ub,Gavela:1980bp}.

In Tab.~\ref{tab:fwidth} we also list the partial decay widths in comparison with the PDG values~\cite{Olive:2016xmw}. Note that the calculated partial decay width of $\Sigma^0\to n \gamma$ is comparable with the other channels although its branching ratio looks very small. This is due to the large total width saturated by the EM transition of $\Sigma^0\to \Lambda \gamma$. The partial width for $\Sigma^0\to n \gamma$ has not been measured in experiment. The measurement of this quantity may be pursued at BESIII with large event samples collected in $J/\psi$ decays.

\begin{table}[ht]
\centering
\caption{The numerical results of the asymmetry parameter are compared with experimental data and other theoretical predictions.}
\begin{ruledtabular}
\begin{tabular}{l|ccccccc}
%\hline \hline
$B_i\to B_f \gamma$   &PDG~\cite{Olive:2016xmw} &Broken SU(3)~\cite{Zenczykowski:2005cs} &ChPT~\cite{Borasoy:1999nt} &Pole model (I)~\cite{Nardulli:1987ub}&Pole model (II)~\cite{Gavela:1980bp} &Our result  \\
\hline
$\Sigma^+\to p\gamma$   &$-0.76\pm0.08$  &$-0.67$&$-0.49$ &$-0.92$ &$-0.80$ &$-0.58\pm0.060$
\\
$\Lambda\to n\gamma$    &-               &$-0.93$&$-0.19$ &$0.91$  &$-0.49$ &$-0.67\pm0.060$
\\
$\Xi^0\to\Lambda\gamma$ &$(-0.70\pm0.07)$&$-0.97$&$0.46$  &$0.07$  &$-0.78$ &$0.72\pm0.11$       \\
$\Xi^0\to\Sigma^0\gamma$&$(-0.69\pm0.06)$&$-0.92$&$0.15$  &$-0.75$ &$-0.96$ &$0.33\pm0.036$
\\
$\Sigma^0\to n\gamma$   &-               &-      &-       &$-0.65$&$-0.98$  &$0.37\pm0.035$\\
%\hline \hline
\end{tabular}
\end{ruledtabular}
\label{tab:as}
\end{table}

\begin{table}[ht]
\centering
\caption{The calculated partial decay widths (in units of $10^{-18}$ GeV) are compared with the experimental data from PDG~\cite{Olive:2016xmw}.}
\begin{ruledtabular}
\begin{tabular}{lccccc}
$B_i\to B_f \gamma$ &$\Sigma^+\to p\gamma$ &$\Lambda\to n\gamma$ &$\Xi^0\to\Lambda\gamma$ &$\Xi^0\to\Sigma^0\gamma$&$\Sigma^0\to n\gamma$ \\
\hline
PDG data~\cite{Olive:2016xmw} &$10.10\pm0.41$ &$4.40\pm0.38$&$2.66\pm0.18$  &$7.56\pm0.33$ & -    \\
Our result                        &$8.73\pm4.88$&$4.59\pm2.42$&$2.20\pm0.73$  &$22.13\pm9.50$&$1.59\pm0.55$    \\
\end{tabular}
\end{ruledtabular}
\label{tab:fwidth}
\end{table}

\section{Summary}
\label{sec:sum}

In this work we revisit the hyperon weak radiative decays in the framework of the NRCQM. The dominance of the pole terms turns out to be crucial for achieving an overall self-consistent description of the available experimental data for hyperon weak radiative decays. This study confirms the nonlocal feature of the hyperon weak radiative transition operators which provides a natural mechanism for evading the Hara theorem, i.e. the PV contributions can come from the intermediate orbital excitations in the NRCQM. We also discuss several interesting selection rules arising from either the EM or the weak interaction vertices. Moreover, there exist cancellations between pole terms that can significantly violate the SU(3) flavor symmetry in the observables. These features suggest nontrivial relations among different hyperon decays. It is interesting to note that the dominance of the pole terms is somehow counterintuitive taking into account the short-ranged property of the weak interactions. Therefore, a coherent study of the hyperon weak radiative decays and confirmation of the dominance of pole terms are crucial for a better understanding of the underlying dynamics. Future studies of possible dynamical effects are strongly recommended. A better description of the transition operators will make the hyperons good probes for long-ranged weak-decay dynamics in nuclear few-body systems.

\begin{acknowledgments}

This work is supported, in part, by the National Natural Science Foundation of China (Grant Nos. 11425525 and 11521505), DFG and NSFC funds to the Sino-German CRC 110 ``Symmetries and the Emergence of Structure in QCD'' (NSFC Grant No. 11261130311), Strategic Priority Research Program of Chinese Academy of Sciences (Grant No. XDB34030302), and National Key Basic Research Program of China under Contract No. 2015CB856700. Q.W. is also supported by the research startup funding at SCNU, Guangdong Provincial funding with Grant No. 2019QN01X172£¬and Science and Technology Program of Guangzhou (No. 2019050001). J.M.R. would like to thank the hospitality provided to him at IHEP, where part of this work was completed, and the support  by the Munich Institute for Astro- and Particle Physics (MIAPP) of the DFG cluster of excellence ``Origin and Structure of the Universe'' during the Workshop ``Deciphering Strong-Interaction Phenomenology through Precision Hadron-Spectroscopy.''

\end{acknowledgments}

\begin{appendix}

\section{Convention}
\label{app:convention}

The following conventions are adopted for the quark and anti-quark field:
\begin{align}
q(x)&=\int \frac{d \bm p}{(2\pi)^{3/2}}\left(\frac{m}{p^0}\right)^{1/2} \sum_s u_s(\bm p)b_s(\bm p)e^{i p\cdot x}+v_s(\bm p)d^\dagger_s(\bm p)e^{-i p\cdot x}, \\
\bar q(x)&=\int\frac{d\bm p}{(2\pi)^{3/2}}\left(\frac{m}{p^0}\right)^{1/2} \sum_s \bar u_s(\bm p)b^\dagger_s(\bm p)e^{-i p\cdot x}+\bar v_s(\bm p)d_s(\bm p)e^{i p\cdot x}.
\end{align}
The commutation and anticommutation relations of the creation and annihilation operators are given by:
\begin{align}
&\{b_s(\bm p),b_{s'}^\dagger(\bm p') \}=\{d_s(\bm p),d_{s'}^\dagger(\bm p') \}=\delta_{ss'}\delta^3(\bm p-\bm p').
\end{align}
The normalization of spinor is $u^\dagger_s(\bm p)u_{s'}(\bm p)=v^\dagger_s(\bm p)v_{s'}(\bm p)=(p^0/m)\delta_{ss'}$. It should be emphasized that the convention of quark field and spinor must match each other in order to keep the nonrelativistic Hamiltonian  independent of any convention.

\section{The Baryon Wave Function Within the Non-relativistic Constituent Quark Model}
\label{app:wave}

We adopt the nonrelativistic quark model wave functions for the baryons~\cite{Isgur:1978xj} in the calculation. The total wave function of hadron consists of four parts: a) the color wave function which is trivial and neglected here; b) the spin wave function; c) the flavor wave function and d) the spatial wave function.

The spin wave functions for a three-quark system are written as:
\begin{align}
&\chi^{s}_{3/2}={}\uparrow \uparrow \uparrow  ,\qquad
&&\chi^{s}_{-1/2}=\frac{1}{\sqrt3}\left( \uparrow\downarrow\downarrow+\downarrow\uparrow\downarrow+\downarrow\downarrow\uparrow\right)  ,\notag\\[-1pt]
&\chi^{s}_{1/2}=\frac{1}{\sqrt3}\left(\uparrow \uparrow \downarrow +\uparrow \downarrow \uparrow +\downarrow \uparrow \uparrow \right)  ,
&&\chi^{s}_{-3/2}={} \downarrow\downarrow\downarrow  ,\\[2pt]
&\chi^{\rho}_{1/2}=\frac{1}{\sqrt2}\left(\uparrow \downarrow \uparrow -\downarrow \uparrow \uparrow \right),
&& \chi^{\lambda}_{1/2}=-\frac{1}{\sqrt6}\left( \uparrow\downarrow\uparrow+\downarrow\uparrow\uparrow-2\uparrow\uparrow\downarrow \right), \notag \\[-1pt]
&\chi^{\rho}_{-1/2}=\frac{1}{\sqrt2}\left(\uparrow \downarrow \downarrow -\downarrow \uparrow \downarrow \right),
&&\chi^{\lambda}_{-1/2}=\frac{1}{\sqrt6}\left( \uparrow\downarrow\downarrow+\downarrow\uparrow\downarrow-2\downarrow\downarrow\uparrow\right) \ ,
\end{align}
where the superscripts, $s$, $a$, $\rho$ and $\lambda$, are used to label the symmetry types of the corresponding wavefunctions, namely overall symmetric and antisymmetric states, and  mixed symmetry states that are either antisymmetric or symmetric under the exchange of the first two quarks, respectively.

The flavor wave functions for the octet baryons are written as~\cite{LeYaouanc:1988fx}:
\begin{align}
&\phi_p^\lambda=\frac{1}{\sqrt{6}}(2 u u d-d u u-u d u),
&&\phi_p^\rho=\frac{1}{\sqrt{2}}(u d u-d u u), \notag \\
&\phi_n^\lambda=\frac{1}{\sqrt{6}}(d u d+u d d-2 d d u),
&&\phi_n^\rho=\frac{1}{\sqrt{2}}(u d d-d u d),\notag \\
&\phi_{\Lambda}^\lambda=\frac{1}{2}(sud+usd-sdu-dsu),
&&\phi_{\Lambda}^\rho=\frac{1}{2\sqrt3}(usd+sdu-sud-dsu-2dus+2uds), \notag \\
&\phi_{\Sigma^+}^\lambda = \frac{1}{\sqrt6}(suu+usu-2uus),
&&\phi_{\Sigma^+}^\rho =\frac{1}{\sqrt2}(suu-usu), \notag \\
&\phi_{\Sigma^0}^\lambda=\frac{1}{2\sqrt3}(sdu+sud+usd+dsu-2uds-2dus),
&&\phi_{\Sigma^0}^\rho=\frac{1}{2}(sud+sdu-usd-dsu), \notag \\
&\phi^\lambda_{\Xi^0}=\frac{1}{\sqrt6}(2ssu-sus-uss),
&&\phi^\rho_{\Xi^0}=\frac{1}{\sqrt2}(sus-uss).
\end{align}

A basis of  spatial wave functions in the momentum space is given by~\cite{LeYaouanc:1988fx}:
\begin{align}
% \Psi_{NLM}(\bm P,\bm p_\rho,\bm p_\lambda)
\Psi^{n_\rho\ell\rho;n_\lambda \ell_\lambda}_{LM}(\bm P,\bm p_\rho,\bm p_\lambda)
=\delta^3(\bm P_{cm}-\bm P)\sum_m \langle l_\rho,m;l_\lambda,M-m|LM \rangle\psi_{n_\rho l_\rho m }(\bm p_\rho)\psi_{n_\lambda l_\lambda M-m }(\bm p_\lambda),
\end{align}
where $\bm P$, $\bm p_\rho$ and $\bm p_\lambda$ are the usual Jacobi coordinates, and
\begin{align}
\psi_{n,l,m}(\bm p)=(i)^l(-1)^n \left[\frac{2n!}{(n+l+1/2)!} \right]^{1/2}\frac{1}{\alpha^{l+3/2}}e^{-\genfrac{}{}{}{1}{\bm p^2}{2\alpha^2}}L_n^{l+1/2}(\bm p^2/\alpha^2)
\mathcal{Y}_{lm}(\bm p).
\end{align}
$n_\rho$ and $n_\lambda$ count the radial excitations and $\bm L$ is the total angular momentum. The $L_n^\nu(x)$ is generalized Laguerre polynomials and $\alpha$ is the oscillator parameter. $ \mathcal{Y}_{lm}(\bm p)$ is solid spherical harmonics.

The baryons are  three-quark systems, and  in the limit of SU(3) flavor symmetry their total wave functions become totally symmetric. With the color wave function total antisymmetric, the rest of part of a total wave function should then be symmetric. The total wave function (except for the color part) of an octet baryon $B$ is written as:
\begin{align}
\label{eq:state5628}
|56,\rep28,0,0,s_z\rangle=\frac{1}{\sqrt2}(\phi_B^\rho \chi^\rho_{s_z}+\phi_B^\lambda \chi^\lambda_{s_z})\Psi^{00;00}_{00}(\bm p_\rho,\bm p_\lambda),
\end{align}
The total wave function of the first orbital excitation states of $[70,\rep28]$ ($N(1535), \ \Lambda(1670), \ \Sigma(1620)$ and $\Xi(1620)$) is written as:
\begin{align}
\label{eq:state7028}
|70,\rep28,1,1,J_z\rangle=\sum_{L_z+S_z=J_z}\langle1 L_z;\frac{1}{2}S_z|\frac12 J_z\rangle \frac{1}{2}\bigl[(\phi_B^\rho \chi_{S_z}^\lambda
+\phi_B^\lambda \chi_{S_z}^\rho)\Psi^{1L_z;00}_{1L_z}(\bm p_\rho,\bm p_\lambda)\notag\\
{}+{}
(\phi_B^\rho \chi_{S_z}^\rho-\phi_B^\lambda \chi_{S_z}^\lambda)\Psi^{0,0;1,L_z}_{1L_z}(\bm p_\rho,\bm p_\lambda) \bigr].
\end{align}
The total wave function of the first orbital excitation states of $[70,\rep48]$ ($N(1650),\Lambda(1800),\Sigma(1750)$ and $\Xi(1690)$) is as follows:
\begin{align}
\label{eq:state7048}
|70,\rep48,1,1,J_z\rangle=\sum_{L_z+S_z=J_z}\langle1 L_z;\frac{3}{2}S_z|\frac12 J_z\rangle \frac{1}{\sqrt2}\left[\phi_B^\rho \chi_{S_z}^s \Psi^{1 L_z;0 0}_{1L_z}(\bm p_\rho,\bm p_\lambda)+
\phi_B^\lambda \chi_{S_z}^s\Psi^{00;1 L_z}_{1L_z}(\bm p_\rho,\bm p_\lambda) \right].
\end{align}
The total wave function of the first orbital excitation states of $[70,\rep{2}{10}]$ ($\Delta(1620)$) is as follows:
\begin{align}
\label{eq:state10210}
|70,\rep{2}{10},1,1,J_z\rangle=\sum_{L_z+S_z=J_z}\langle1 L_z;\frac{1}{2}S_z|\frac12 J_z\rangle \frac{1}{\sqrt2}\left[\phi_B^s \chi_{S_z}^\rho \Psi^{1 L_z;0 0}_{1L_z}(\bm p_\rho,\bm p_\lambda)+
\phi_B^s \chi_{S_z}^\lambda\Psi^{00;1 L_z}_{1L_z}(\bm p_\rho,\bm p_\lambda) \right].
\end{align}
The total wave function of the first orbital excitation state of $[70, \ ^21]$ ($\Lambda(1405)$) is
\begin{align}
\label{eq:state7021}
|70,\rep21,1,1,J_z\rangle=\sum_{L_z+S_z=J_z}\langle1 L_z;\frac{3}{2}S_z|\frac12 J_z\rangle \frac{1}{\sqrt2}\left[\phi_\Lambda^a \chi_{S_z}^\lambda \Psi^{1L_z;00}_{1L_z}(\bm p_\rho,\bm p_\lambda)-
\phi_\Lambda^a \chi_{S_z}^\lambda\Psi^{00;1 L_z}_{1L_z}(\bm p_\rho,\bm p_\lambda) \right],
\end{align}
where
\begin{align}
\phi_\Lambda^a=\frac{1}{\sqrt6}(uds+ dsu+sud-dus-usd -sdu).
\end{align}

\section{Amplitudes}
\label{app:amp}
The transition amplitudes with the EM operator $H_{EM}^+$ are given in this section. Note that only the nonzero amplitudes are listed. The nonzero amplitudes are labeled as $\Mc_{PC/PV,A/B}(B)$,  where the spin indexes $\epsilon$, $S_z^f=+1/2$ and $S_z^i=-1/2$ are omitted. $B$ in the parentheses is the name of intermediate baryon. In the following part, $S_F(B)$ stands for the propagator of the intermediate baryon and is written as:
\begin{align}
S_F(B)=\frac{i(\slashed p_B+M_{B})}{ p^2_{B} -M^2_{B}+i\Gamma_{B}M_{B}}\approx\frac{2 i M_{B}}{ p^2_{B} -M^2_{B}+i\Gamma_{B}M_{B}},
\end{align}
where $M_B \ (\Gamma_B)$ is the mass (width) of the baryon and $p_B$ its four momentum.

\begin{itemize}
\item $\Sigma^+\to p \gamma$
\begin{align}
\Mc_{PC,A}(p)&=
\left[\frac{3 \alpha ^3 G_F V_{\text{ud}} V_{\text{us}}}{2 \pi ^{3/2}} \right]
S_F(p)
\left[ -\frac{\eq k}{4 \sqrt{2} \pi ^{3/2} \sqrt{k^0} m_q} \right] e^{-\genfrac{}{}{}{1}{k^2}{6 \alpha ^2}},
\\
\Mc_{PV,A}(N(1535))&=\left[-\frac{3 i \alpha ^4 G_F V_{\text{ud}} V_{\text{us}}}{2 \sqrt{2} \pi ^{3/2} m_q}  \right]
S_F(N(1535))
\left[\frac{i \eq  \left(2 \alpha ^2-k^2\right)}{24 \pi ^{3/2} \alpha  \sqrt{k^0} m_q} \right]e^{-\genfrac{}{}{}{1}{k^2}{6 \alpha ^2}}.
\end{align}

\begin{align}
\Mc_{PC,B}(\Sigma^+)&=e^{-\genfrac{}{}{}{1}{k^2}{6 \alpha ^2}}\left[-\frac{\eq k }{4 \sqrt{2} \pi ^{3/2} \sqrt{k^0} m_q} \right]
S_F(\Sigma^+)
\left[ \frac{3 \alpha ^3 G_F V_{\text{ud}} V_{\text{us}}}{2 \pi ^{3/2}} \right],
\\
\Mc_{PV,B}(\Sigma(1620))&=e^{-\genfrac{}{}{}{1}{k^2}{6 \alpha ^2}} \left[ -\frac{i \eq \left(2 \alpha ^2-k^2\right)}{24 \pi ^{3/2} \alpha  \sqrt{k^0} m_q} \right]
 S_F(\Sigma(1620))
\left[ \frac{3 i \alpha ^4 G_F V_{\text{ud}} V_{\text{us}}}{2 \sqrt{2} \pi ^{3/2} m_q} \right].
\end{align}

\item $\Lambda \to n \gamma$
\begin{align}
\Mc_{PC,A}(n)&=\left[-\frac{\sqrt{\frac{3}{2}} \alpha ^3 G_F V_{\text{ud}} V_{\text{us}}}{2 \pi ^{3/2}} \right]
S_F(n)
\left[ \frac{\eq k }{6 \sqrt{2} \pi ^{3/2} \sqrt{k^0} m_q} \right]e^{-\genfrac{}{}{}{1}{k^2}{6 \alpha ^2}},
\\
\Mc_{PV,A}(N(1535))&=\left[\frac{i \sqrt{3} \alpha ^4 G_F V_{\text{ud}} V_{\text{us}}}{4 \pi ^{3/2} m_q} \right]
S_F(N(1535))
\left[-\frac{i \eq  \left(6 \alpha ^2-k^2\right)}{72 \pi ^{3/2} \alpha  \sqrt{k^0} m_q} \right]e^{-\genfrac{}{}{}{1}{k^2}{6 \alpha ^2}},
\\
\Mc_{PV,A}(N(1650))&=\left[\frac{i \sqrt{3} \alpha ^4 G_F V_{\text{ud}} V_{\text{us}}}{2 \pi ^{3/2} m_q} \right]
S_F(N(1650))
\left[\frac{i \eq k^2 }{72 \pi ^{3/2} \alpha  \sqrt{k^0} m_q} \right]e^{-\genfrac{}{}{}{1}{k^2}{6 \alpha ^2}}.
\end{align}

\begin{align}
\Mc_{PC,B}(\Lambda)&=e^{-\genfrac{}{}{}{1}{k^2}{6 \alpha ^2}}\left[\frac{\eq k }{12 \sqrt{2} \pi ^{3/2} \sqrt{k^0} m_q} \right]
S_F(\Lambda)
\left[-\frac{\sqrt{\frac{3}{2}} \alpha ^3 G_F V_{\text{ud}} V_{\text{us}}}{2 \pi ^{3/2}} \right],
\\
\Mc_{PC,B}(\Sigma^0)&=e^{-\genfrac{}{}{}{1}{k^2}{6 \alpha ^2}}\left[-\frac{\eq k }{4 \sqrt{6} \pi ^{3/2} \sqrt{k^0} m_q} \right]
S_F(\Sigma^0)
\left[\frac{3 \alpha ^3 G_F V_{\text{ud}} V_{\text{us}}}{2 \sqrt{2} \pi ^{3/2}} \right],
\\
\Mc_{PV,B}(\Lambda(1670))&=e^{-\genfrac{}{}{}{1}{k^2}{6 \alpha ^2}}\left[\frac{i\eq  \left(6 \alpha ^2-k^2\right)}{144 \pi ^{3/2} \alpha  \sqrt{k^0} m_q} \right]
S_F(\Lambda(1670))
\left[-\frac{i \sqrt{3} \alpha ^4 G_F V_{\text{ud}} V_{\text{us}}}{4 \pi ^{3/2} m_q} \right],
\\
\Mc_{PV,B}(\Lambda(1800))&=e^{-\genfrac{}{}{}{1}{k^2}{6 \alpha ^2}}\left[-\frac{i \eq k^2 }{144 \pi ^{3/2} \alpha  \sqrt{k^0} m_q} \right]
S_F(\Lambda(1800))
\left[-\frac{i \sqrt{3} \alpha ^4 G_F V_{\text{ud}} V_{\text{us}}}{2 \pi ^{3/2} m_q} \right],
\\
\Mc_{PV,B}(\Lambda(1450))&=e^{-\genfrac{}{}{}{1}{k^2}{6 \alpha ^2}}\left[-\frac{i \eq  \left(2 \alpha ^2-k^2\right)}{48 \pi ^{3/2} \alpha  \sqrt{k^0} m_q} \right]
S_F(\Lambda(1450))
\left[\frac{i \sqrt{3} \alpha ^4 G_F V_{\text{ud}} V_{\text{us}}}{2 \pi ^{3/2} m_q} \right],
\\
\Mc_{PV,B}(\Sigma(1620))&=e^{-\genfrac{}{}{}{1}{k^2}{6 \alpha ^2}}\left[ -\frac{i \eq  \left(6 \alpha ^2-k^2\right)}{48 \sqrt{3} \pi ^{3/2} \alpha  \sqrt{k^0} m_q}\right]
S_F(\Sigma(1620))
\left[\frac{3 i \alpha ^4 G_F V_{\text{ud}} V_{\text{us}}}{4 \pi ^{3/2} m_q} \right],
\\
\Mc_{PV,B}(\Sigma(1750))&=e^{-\genfrac{}{}{}{1}{k^2}{6 \alpha ^2}}\left[ \frac{i\eq k^2 }{48 \sqrt{3} \pi ^{3/2} \alpha  \sqrt{k^0} m_q} \right] S_F(\Sigma(1750))
\left[ \frac{3 i \alpha ^4 G_F V_{\text{ud}} V_{\text{us}}}{2 \pi ^{3/2} m_q}\right].
\end{align}

\item $\Xi^0\to\Lambda\gamma$
\begin{align}
\Mc_{PC,A}(\Lambda)&=\left[ \frac{2 \sqrt{3} \alpha ^3 \beta ^3 G_F V_{\text{ud}} V_{\text{us}}}{\pi ^{3/2} \left(\alpha ^2+\beta ^2\right)^{3/2}} \right] S_F(\Lambda)
\left[\frac{\eq k }{12 \sqrt{2} \pi ^{3/2} \sqrt{k^0} m_q} \right]e^{-\genfrac{}{}{}{1}{k^2}{6 \alpha ^2}},
\\
\Mc_{PV,A}(\Lambda(1670))&=\left[-\frac{i \sqrt{6} \alpha ^4 \beta ^3 G_F V_{\text{ud}} V_{\text{us}}}{\pi ^{3/2} \left(\alpha ^2+\beta ^2\right)^{3/2} m_q} \right]
S_F(\Lambda(1670))
\left[-\frac{i \eq  \left(6 \alpha ^2-k^2\right)}{144 \pi ^{3/2} \alpha  \sqrt{k^0} m_q} \right]e^{-\genfrac{}{}{}{1}{k^2}{6 \alpha ^2}},
\\
\Mc_{PV,A}(\Lambda(1800))&=\left[ -\frac{2 i \sqrt{6} \alpha ^4 \beta ^3 G_F V_{\text{ud}} V_{\text{us}}}{\pi ^{3/2} \left(\alpha ^2+\beta ^2\right)^{3/2} m_q} \right]
S_F(\Lambda(1800))
\left[ \frac{i \eq k^2 }{144 \pi ^{3/2} \alpha  \sqrt{k^0} m_q} \right]e^{-\genfrac{}{}{}{1}{k^2}{6 \alpha ^2}},
\\
\Mc_{PV,A}(\Lambda(1450))&=\left[ -\frac{i \sqrt{6} \alpha ^4 \beta ^3 G_F V_{\text{ud}} V_{\text{us}}}{\pi ^{3/2} \left(\alpha ^2+\beta ^2\right)^{3/2} m_q} \right]
S_F(\Lambda(1450))
\left[ \frac{i \eq  \left(2 \alpha ^2-k^2\right)}{48 \pi ^{3/2} \alpha  \sqrt{k^0} m_q}\right]e^{-\genfrac{}{}{}{1}{k^2}{6 \alpha ^2}}.
\end{align}

\begin{align}
\Mc_{PC,B}(\Xi^0)&=e^{-\genfrac{}{}{}{1}{k^2}{6 \beta ^2}}\left[\frac{\eq k }{6 \sqrt{2} \pi ^{3/2} \sqrt{k^0} m_q} \right]
S_F(\Xi^0)
\left[ \frac{2 \sqrt{3} \alpha ^3 \beta ^3 G_F V_{\text{ud}} V_{\text{us}}}{\pi ^{3/2} \left(\alpha ^2+\beta ^2\right)^{3/2}} \right],
\\
\Mc_{PV,B}(\Xi(1620))&=e^{-\genfrac{}{}{}{1}{k^2}{6 \beta ^2}}\left[\frac{i \eq  \left(6 \beta ^2-k^2\right)}{72 \pi ^{3/2} \beta  \sqrt{k^0} m_q} \right]
S_F(\Xi(1620))
\left[ \frac{i \sqrt{6} \alpha ^3 \beta ^4 G_F V_{\text{ud}} V_{\text{us}}}{\pi ^{3/2} \left(\alpha ^2+\beta ^2\right)^{3/2} m_q} \right],
\\
\Mc_{PV,B}(\Xi(1690))&=e^{-\genfrac{}{}{}{1}{k^2}{6 \beta ^2}}\left[ -\frac{i \eq k^2 }{72 \pi ^{3/2} \beta  \sqrt{k^0} m_q} \right] S_F(\Xi(1690))
\left[ \frac{2 i \sqrt{6} \alpha ^3 \beta ^4 G_F V_{\text{ud}} V_{\text{us}}}{\pi ^{3/2} \left(\alpha ^2+\beta ^2\right)^{3/2} m_q} \right].
\end{align}

\item $\Xi^0\to\Sigma^0\gamma$
\begin{align}
\Mc_{PC,A}(\Lambda)&=\left[ \frac{2 \sqrt{3} \alpha ^3 \beta ^3 G_F V_{\text{ud}} V_{\text{us}}}{\pi ^{3/2} \left(\alpha ^2+\beta ^2\right)^{3/2}} \right] S_F(\Lambda)
\left[ -\frac{\eq k }{4 \sqrt{6} \pi ^{3/2} \sqrt{k^0} m_q} \right]e^{-\genfrac{}{}{}{1}{k^2}{6 \alpha ^2}},
\\
\Mc_{PV,A}(\Lambda(1670))&=\left[ -\frac{i \sqrt{6} \alpha ^4 \beta ^3 G_F V_{\text{ud}} V_{\text{us}}}{\pi ^{3/2} \left(\alpha ^2+\beta ^2\right)^{3/2} m_q} \right]
S_F(\Lambda(1670))
\left[ \frac{i \eq  \left(6 \alpha ^2-k^2\right)}{48 \sqrt{3} \pi ^{3/2} \alpha  \sqrt{k^0} m_q} \right]e^{-\genfrac{}{}{}{1}{k^2}{6 \alpha ^2}},
\\
\Mc_{PV,A}(\Lambda(1800))&=\left[ -\frac{2 i \sqrt{6} \alpha ^4 \beta ^3 G_F V_{\text{ud}} V_{\text{us}}}{\pi ^{3/2} \left(\alpha ^2+\beta ^2\right)^{3/2} m_q} \right]
S_F(\Lambda(1800))
\left[ -\frac{i \eq k^2 }{48 \sqrt{3} \pi ^{3/2} \alpha  \sqrt{k^0} m_q} \right]e^{-\genfrac{}{}{}{1}{k^2}{6 \alpha ^2}},
\\
\Mc_{PV,A}(\Lambda(1450))&=\left[ -\frac{i \sqrt{6} \alpha ^4 \beta ^3 G_F V_{\text{ud}} V_{\text{us}}}{\pi ^{3/2} \left(\alpha ^2+\beta ^2\right)^{3/2} m_q} \right]
S_F(\Lambda(1450))
\left[ \frac{i \eq  \left(2 \alpha ^2-k^2\right)}{16 \sqrt{3} \pi ^{3/2} \alpha  \sqrt{k^0} m_q} \right]e^{-\genfrac{}{}{}{1}{k^2}{6 \alpha ^2}}.
\end{align}

\item $\Sigma^0\to n \gamma$
\begin{align}
\Mc_{PC,A}(n)&=\left[ \frac{3 \alpha ^3 G_F V_{\text{ud}} V_{\text{us}}}{2 \sqrt{2} \pi ^{3/2}} \right]
S_F(n)
\left[ \frac{\eq k }{6 \sqrt{2} \pi ^{3/2} \sqrt{k^0} m_q} \right]e^{-\genfrac{}{}{}{1}{k^2}{6 \alpha ^2}},
\\
\Mc_{PV,A}(N(1535))&=\left[ -\frac{3 i \alpha ^4 G_F V_{\text{ud}} V_{\text{us}}}{4 \pi ^{3/2} m_q} \right]
S_F(N(1535))
\left[ -\frac{i \eq  \left(6 \alpha ^2-k^2\right)}{72 \pi ^{3/2} \alpha  \sqrt{k^0} m_q} \right]e^{-\genfrac{}{}{}{1}{k^2}{6 \alpha ^2}},
\\
\Mc_{PV,A}(N(1650))&=\left[ -\frac{3 i \alpha ^4 G_F V_{\text{ud}} V_{\text{us}}}{2 \pi ^{3/2} m_q} \right]
S_F(N(1650))
\left[ \frac{i \eq k^2 }{72 \pi ^{3/2} \alpha  \sqrt{k^0} m_q} \right]e^{-\genfrac{}{}{}{1}{k^2}{6 \alpha ^2}}.
\end{align}

\begin{align}
\Mc_{PC,B}(\Lambda)&=e^{-\genfrac{}{}{}{1}{k^2}{6 \alpha ^2}}\left[ -\frac{\eq k }{4 \sqrt{6} \pi ^{3/2} \sqrt{k^0} m_q} \right]
S_F(\Lambda)
\left[-\frac{\sqrt{\frac{3}{2}} \alpha ^3 G_F V_{\text{ud}} V_{\text{us}}}{2 \pi ^{3/2}} \right],
\\
\Mc_{PC,B}(\Sigma^0)&=e^{-\genfrac{}{}{}{1}{k^2}{6 \alpha ^2}}\left[ -\frac{\eq k }{12 \sqrt{2} \pi ^{3/2} \sqrt{k^0} m_q}\right]
S_F(\Sigma^0)
\left[\frac{3 \alpha ^3 G_F V_{\text{ud}} V_{\text{us}}}{2 \sqrt{2} \pi ^{3/2}} \right],
\\
\Mc_{PV,B}(\Lambda(1670))&=e^{-\genfrac{}{}{}{1}{k^2}{6 \alpha ^2}}\left[ -\frac{i \eq  \left(6 \alpha ^2-k^2\right)}{48 \sqrt{3} \pi ^{3/2} \alpha  \sqrt{k^0} m_q} \right]
 S_F(\Lambda(1670))
\left[ -\frac{i \sqrt{3} \alpha ^4 G_F V_{\text{ud}} V_{\text{us}}}{4 \pi ^{3/2} m_q} \right],
\\
\Mc_{PV,B}(\Lambda(1800))&=e^{-\genfrac{}{}{}{1}{k^2}{6 \alpha ^2}}\left[ \frac{i \eq k^2 }{48 \sqrt{3} \pi ^{3/2} \alpha  \sqrt{k^0} m_q} \right]
S_F(\Lambda(1800))
\left[ -\frac{i \sqrt{3} \alpha ^4 G_F V_{\text{ud}} V_{\text{us}}}{2 \pi ^{3/2} m_q} \right],
\\
\Mc_{PV,B}(\Lambda(1450))&=e^{-\genfrac{}{}{}{1}{k^2}{6 \alpha ^2}}\left[ -\frac{i \eq  \left(2 \alpha ^2-k^2\right)}{16 \sqrt{3} \pi ^{3/2} \alpha  \sqrt{k^0} m_q} \right]
S_F(\Lambda(1450))
\left[ \frac{i \sqrt{3} \alpha ^4 G_F V_{\text{ud}} V_{\text{us}}}{2 \pi ^{3/2} m_q} \right],
\\
\Mc_{PV,B}(\Sigma(1620))&=e^{-\genfrac{}{}{}{1}{k^2}{6 \alpha ^2}}\left[ -\frac{i \eq  \left(6 \alpha ^2-k^2\right)}{144 \pi ^{3/2} \alpha  \sqrt{k^0} m_q} \right]
S_F(\Sigma(1620))
\left[ \frac{3 i \alpha ^4 G_F V_{\text{ud}} V_{\text{us}}}{4 \pi ^{3/2} m_q} \right],
\\
\Mc_{PV,B}(\Sigma(1750))&=e^{-\genfrac{}{}{}{1}{k^2}{6 \alpha ^2}}\left[ \frac{i \eq k^2 }{144 \pi ^{3/2} \alpha  \sqrt{k^0} m_q} \right] S_F(\Sigma(1750))
\left[ \frac{3 i \alpha ^4 G_F V_{\text{ud}} V_{\text{us}}}{2 \pi ^{3/2} m_q} \right].
\end{align}
\end{itemize}

\end{appendix}

\end{document}